\documentclass{article}
\usepackage{spconf,amsmath,graphicx}
\usepackage[ruled,linesnumbered]{algorithm2e}
\usepackage{subfig}
\usepackage{amsfonts}
\usepackage{multirow}

\usepackage{bm}
\usepackage{txfonts}
\usepackage{amssymb}
\usepackage{array}
\usepackage{cases}
\usepackage{algpseudocode}
\usepackage{enumerate}
\usepackage{color,xcolor}
\usepackage[justification=centering]{caption}

\title{Stein Variational Gradient Descent-based Detection for Random Access with Preambles in MTC}
\name{Xin Zhu$^*$ \quad Hongyi Pan$^\dagger$  \quad Salih Atici$^*$ \quad Ahmet Enis Cetin$^*$\thanks{This work was supported by NSF IDEAL 2217023.}}
\address{$^*$Department of Electrical and Computer Engineering, University of Illinois Chicago, USA\\
$^\dagger$Machine \& Hybrid Intelligence Lab, Northwestern University, USA}
\begin{document}
%
\maketitle\small
\begin{abstract}
Traditional preamble detection algorithms have low accuracy in the grant-based random access scheme in massive machine-type communication (mMTC). We present a novel preamble detection algorithm based on Stein variational gradient descent (SVGD) at the second step of the random access procedure. It efficiently leverages deterministic updates of particles for continuous inference. To further enhance the performance of the SVGD detector, especially in a dense user scenario, we propose a normalized SVGD detector with momentum. It utilizes the momentum and a bias correction term to reduce the preamble estimation errors during the gradient descent process. Simulation results show that the proposed algorithm performs better than Markov Chain Monte Carlo-based approaches in terms of detection accuracy.
\end{abstract}
\begin{keywords}
Preamble detection, Stein variational gradient descent, grant-based random access, massive machine-type communication (mMTC).
\end{keywords}
\section{Introduction}
\label{sec:intro}
Under the background of massive machine-type communication (mMTC)~\cite{hsu2023hyper,ma2023model}, the future wireless communication system needs to hold a large number of devices. However, the preamble resources are limited. With the number of devices increasing, the preamble collision problem~\cite{ye2023density,choi2021grant,yin2023learning} becomes more serious, which increases the difficulty of the communication systems design. To reduce power consumption and achieve massive connection in the future Internet of Things (IoT) scenarios \cite{sarker2023internet}, communication systems need a random access scheme that can alleviate the preamble collision problem. At present, such random access schemes are mainly divided into two categories: grant-based random access (GBRA) and grant-free random access (GFRA)~\cite{che2023massive,zhang2020hybrid,liu2023grant}.

In this paper, we use the GBRA scheme. GBRA \cite{hasan2013random,reddy2021successive,bezerra2018rach,zhou2015low} includes a four-step handshake random access protocol: (1) Each active user randomly selects a preamble from the preamble pool. (2) The base station (BS) delivers random access responses to users.
(3) Users return MSG3 to the BS. (4) The BS performs random access conflict resolution. In the dense user scenario, a large number of users access the BS at the same time. Therefore, preamble collision can not be avoided \cite{jang2016early}, which significantly increases uplink transmission signaling overhead. In addition, it also increases the access delay and reduces the throughput of the system~\cite{han2020grant}. In the GBRA, colliding users can only be recognized by the base station in the final step. However, the base station will still allocate time-frequency resources to the colliding users in the second step, which causes a waste of resources. Hence, efficient schemes are necessary for random access to detect preamble collisions in advance.

A preamble detection random access scheme based on Markov Chain Monte Carlo (MCMC) was proposed in \cite{choi2018mcmc}. This scheme establishes the maximum a posteriori (MAP) estimation model~\cite{gauvain1994maximum} to detect preambles. However, the MAP estimation needs the prior distribution of the estimated variables. Additionally, the MCMC method increases the diversity of particles through the randomness of sampling, which decreases the accuracy of preamble detection.

To detect preamble collision early and improve the utilization of wireless resources, we establish a preamble detection model based on maximum likelihood estimation~\cite{alma9977926012005897} without using any prior distribution in the second step of handshaking. Furthermore, we propose two efficient algorithms based on Stein variational gradient descent (SVGD) \cite{liu2016stein,liu2016kernelized,zhao2023stein,nusken2023geometry} to find an approximate solution to the maximum likelihood estimation problem. 

The contributions are summarized as follows: (1) We propose a maximum likelihood estimation model based on the SVGD detector to detect preambles in the dense user scenario. (2) Through error analysis, we propose the normalized SVGD (NSVGD) detector with momentum. It has better robustness and a higher preamble detection accuracy than the SVGD detector and MCMC-based methods.

\section{Background}

\subsection{Stein Variational Gradient Descent (SVGD)}

SVGD is a particle-based variational inference algorithm~\cite{liu2016stein}. It efficiently utilizes gradient information for approximating the target distribution through deterministic updates of particle methods. Firstly, $n$ particles are generated by the uniform distribution or some other distribution.
Then, the particles are updated using an optimized gradient ${\varphi}$. Liu~\textit{et al.}~\cite{liu2016stein} proved that the perturbation direction given by ${\varphi}$ is optimal because it corresponds to the steepest descent on the Kullback–Leibler divergence. After sufficient iterations, the obtained particles follow the target distribution. In this paper, we will apply the SVGD algorithm to take samples from the complicated distribution.
	
		
\subsection{System Model}
Assume that there are $N$ active users in the cell. Each user is equipped with an antenna and the BS is equipped with $K$ antennas. The number of preambles is $M$, and the length of each preamble is $S$. Each active user randomly selects a preamble in the pool. Then, the signal received by the BS through antenna $j$ can be expressed as:
\begin{equation}\label{Eq:yph}
		\mathbf{y}_{j}=\sum_{i=1}^{N}\mathbf p_{(i)}H_{i,j}e_{i}+\mathbf n_{j},
\end{equation}
for $j=1,\dots,K$, where $\mathbf n_{j}\sim \mathcal{CN}(0,\beta\mathbf I)$ is the background noise at the $j$-th antenna. $\beta$ is the noise power. $\mathcal{CN}(\mathbf d,\mathbf F)$ denotes the distribution of circularly symmetric complex Gaussian (CSCG) random vectors. 
$\mathbf d$ and $\mathbf F$ are the mean vector and covariance matrix respectively.
$H_ {i,j} $ is the channel coefficient from $i$-th active device to $j$-th antenna. $e_ {i} $ indicates the data symbol sent by the active user to the base station. $ \mathbf p_ {(i)}\in\mathbb{C}^S $ stands for the preamble sequence selected by $i$-th active user.

Suppose $x_{m}$ indicates the number of active users who select the $m$-th preamble. The numbers of users selecting each preamble are represented by a vector $\mathbf x =[x_{1},\dots,x_{m},\dots,x_{M}]^\mathrm{T}$, where $x_{m}\in[0, N]$.
Since our model is designed to detect preamble collision as mentioned in Section \ref{sec:intro}, we focus on estimating $\mathbf x $ in the following steps: Firstly, we define the likelihood of $\mathbf x $:
   \begin{equation}\label{Eq:Lik}
   		{\rm {Lik}(\mathbf x)}=f({\mathbf y_{j}}\mid \mathbf x),
   \end{equation}
where $f({\mathbf y_{j}}\mid \mathbf x)$ represents the Likelihood function of $\mathbf x$ given for $\mathbf y_{j}$.

Then, let $\mathbf w_{j}=[w_{1,k},w_{2,j},\dots,w_{m,j},\dots,w_{M,j}]^\mathrm{T}$, and
\begin{equation}\label{200}
	w_{m,j}=\sum_{i\in\mathcal{\mathcal{N}}_m}H_{i,j}e_{i},
\end{equation}
where $\mathcal{N}_m$ stands for the index set of active users who select the $m$-th preamble.
We assume that the signal transmitted by the active device experiences independent Rayleigh fading:
\begin{equation}\label{201}
		H_{i,j}\sim \mathcal{CN}(0,\delta^2)
\end{equation}

Next, let $\mathbf P = [\mathbf p_{1}, \dots, \mathbf p_{M}] $. The Eq.~(\ref{Eq:yph}) can be rewritten as:
\begin{equation}\label{202}
		\mathbf y_{j}=\mathbf P\mathbf w_{j}+\mathbf n_{j},
\end{equation}

According to Eq. (\ref{200}) and Eq. (\ref{201}), $\mathbf w_{j}$ is a CSCG vector for a given $\mathbf x$. Furthermore, from Eq. (\ref{201}) and Eq. (\ref{202}), $\mathbf y_{j}$ is also a CSCG vector for a given $\mathbf x$. Its mean value is 0 and its covariance matrix is:
\begin{equation}\label{2}
		\mathbb{E}[\mathbf y_{j}\mathbf y_{j}^{\mathrm{H}} \mid \mathbf x]=\delta^2\mathbf P\mathbf V_{\mathbf x}\mathbf P^{\mathrm{H}}+\beta\mathbf I=\phi(\mathbf{x}),
\end{equation}
where $\mathbf V_{\mathbf x}$ = diag$(x_{1}\dots x_{M})$. $(\cdot)^\mathrm{H}$ represents conjugate transpose. 
According to Eq. (\ref{2}), we have
\begin{equation}\label{3}
		\mathbf  y_{j}\mid\mathbf  x\sim CN(\mathbf 0,\phi(\mathbf{x})),
\end{equation}
\begin{equation}\label{Eq:lnfyx}
\begin{split}
		\ln f(\mathbf y_{j}\mid\mathbf x)		
		=-\mathbf y_{j}^{\mathrm{H}}(\phi(\mathbf{x}))^{-1}\mathbf y_{j}-\ln\det(\phi(\mathbf{x}))+\xi,
\end{split}
\end{equation}
where $\xi$ represents a constant. $\det(\cdot)$ stands for the determinant of the matrix.
According to $f(\{\mathbf y_{j}\}\mid\mathbf x)=\prod\nolimits_{j=1}^K f(\mathbf y_{j}\mid\mathbf x)$, log-likelihood function$\ln f(\{\mathbf y_{j}\}\mid\mathbf x)	$ can be represented as:
\begin{equation}\label{Eq:lnall}
\begin{split}
		\ln f(\{\mathbf y_{j}\}\mid\mathbf x)			
		=\sum\limits_{j=1}^{K}\ln f(\mathbf y_{j}\mid\mathbf x),
\end{split}
\end{equation}

Finally, the maximum likelihood estimation model can be derived by using the log-likelihood function:
\begin{equation}\label{4}
			\tilde{\mathbf x}=\mathop{\arg\max}f(\{\mathbf y_{j}\}\mid\mathbf x).
\end{equation}

The computational complexity can be expressed as $(N+1)^M$, which increases exponentially with $M$. Therefore, the computational complexity of the maximum likelihood detection increases significantly when $M$ becomes large.

\section{Methodology}
Since maximum likelihood detection is computationally prohibitive, the SVGD is employed to find an approximate solution. Specifically, we use SVGD to take samples of $\mathbf x$ from the distribution with the density function $q(\mathbf x)=f(\{\mathbf y_{j}\}\mid\mathbf x)$.
  
\subsection{SVGD Detector}
\begin{figure*}[!htbp]
	\centering
	\includegraphics[scale=0.47]{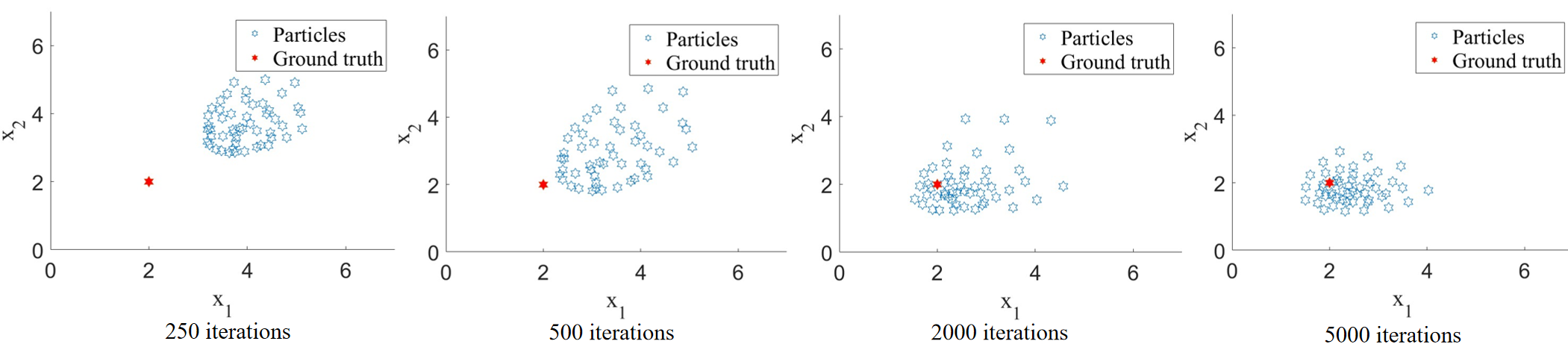}
	\caption{Updating process of particles}
	\label{Fig: paticles}
\end{figure*}

Eq.~(\ref{Eq:lnfyx}) shows that for any $\mathbf x$, $f(\mathbf y_{j}\mid\mathbf x)\ge 0$, which satisfies the SVGD requirement that the density function is positive~\cite{liu2016stein}. Then, we approximate $q(\mathbf x)$ with a set of particles $\{\mathbf x_{i}\}_{i=1}^n$, where $\mathbf x_{i}\in\mathbb{R}^{M}$ and $n$ is the number of particles.
Next, $\{\mathbf x_{i}^0\}_{i=1}^n$ is used to initialize particles and SVGD  updates the particles iteratively by 
\begin{equation}\label{iter}
		\mathbf x_{i}^{t+1}\leftarrow \mathbf x_{i}^{t}+\varepsilon{\varphi}(\mathbf x_i^{t}),
\end{equation}
where $\varepsilon$ is a step size. $\varphi(\cdot)$ is a velocity field that drives the distribution of particles toward the target. It is computed as: 
 \begin{equation}\label{Eq:gradient}
 {\varphi}(\mathbf x)=\frac{1}{n}\sum_{l=1}^{n}[k(\mathbf x_l^{t},\mathbf x)\bigtriangledown_{\mathbf x_l^{t}}\log q(\mathbf x_l^{t})+\bigtriangledown_{\mathbf x_l^{t}}k(\mathbf x_l^{t},\mathbf x)],
 \end{equation}
where $k(\mathbf x_l^{t},\mathbf x_i^{t})$ is the kernel function. The selection of $k(\mathbf x_l^{t},\mathbf x_i^{t})$ will be introduced in section~\ref{simulation}. The cost of computing Eq. (\ref{Eq:gradient}) requires $O(nMKS^2)$ multiplications.
After a number of iterations, we obtain a set of points $\{\mathbf {\tilde{x}}_{i}\}_{i=1}^n$ that approximate our target distribution with the density function $q(\mathbf x)$, so we can estimate the number of users selecting each preamble using $\{\mathbf {\tilde{x}}_{i}\}_{i=1}^n$. 
For example, as shown in Fig.~\ref{Fig: paticles}, given that $M=2, S=2, N=4, K=20, n=50$, the purpose is to estimate $\mathbf{x}=[x_1,x_2]$. At the beginning, fifty particles are initialized with a random distribution $q_0(\mathbf{x})$. Then, the particles are updated using Eq.~(\ref{Eq:gradient}). 
Repeating this iteration, a path of distributions $\{\mathbf x_{i}\}_{i=1}^{50}$ between the $q_0(\mathbf{x})$ and $q(\mathbf{x})$ is constructed. Finally, the particles converge to the ground truth. 

\subsection{Error Analysis of the SVGD Detector}	
		
Given a noisy environment, there will be a large noisy power $\beta$.  
If $\beta$ is significantly larger than each entry of matrix $\psi(\mathbf{x})=\delta^2\mathbf P\mathbf V_{\mathbf x}\mathbf P^{\mathrm{H}}$, $\phi(\mathbf{x})=\delta^2\mathbf P\mathbf V_{\mathbf x}\mathbf P^{\mathrm{H}}+\beta\mathbf I$ will be independent to the change of $\psi(\mathbf{x})$. Then, $\phi(\mathbf{x}) \thickapprox \beta\mathbf I$. 
Therefore, Eq.~(\ref{Eq:lnall}) can be computed as:
\begin{equation}\label{Lc}
		\log q(\mathbf x_i^{t})
		=-\sum_{j=1}^{K}\mathbf y_j^{\mathrm{H}}\mathbf (\beta\mathbf I)^{-1}\mathbf y_{j}-K\ln\det(\beta\mathbf I)+\xi,	
\end{equation}
thus, $\bigtriangledown_{\mathbf{x}_i^{t}}\log q(\mathbf x_i^{t})=0$. Then ${\varphi}(\mathbf x)$ can be computed as:
\begin{equation}\label{Eq:GL}
		{\varphi}(\mathbf x)=\frac{1}{n}\sum_{i=1}^{n}[\bigtriangledown_{\mathbf x_i^{t}}k(\mathbf x_i^{t},\mathbf x)],
\end{equation}
from Eq.~(\ref{Eq:GL}), only $\bigtriangledown_{\mathbf x_i^{t}}k(\mathbf x_i^{t},\mathbf x)$ can be used to update particles. However, $\bigtriangledown_{\mathbf x_i^{t}}k(\mathbf x_i^{t},\mathbf x)$ does not include any information about the target density function $q(\mathbf x)$. As a consequence, errors are introduced during the update process. 

\begin{figure}[htbp]
	\centering
	\includegraphics[scale=0.13]{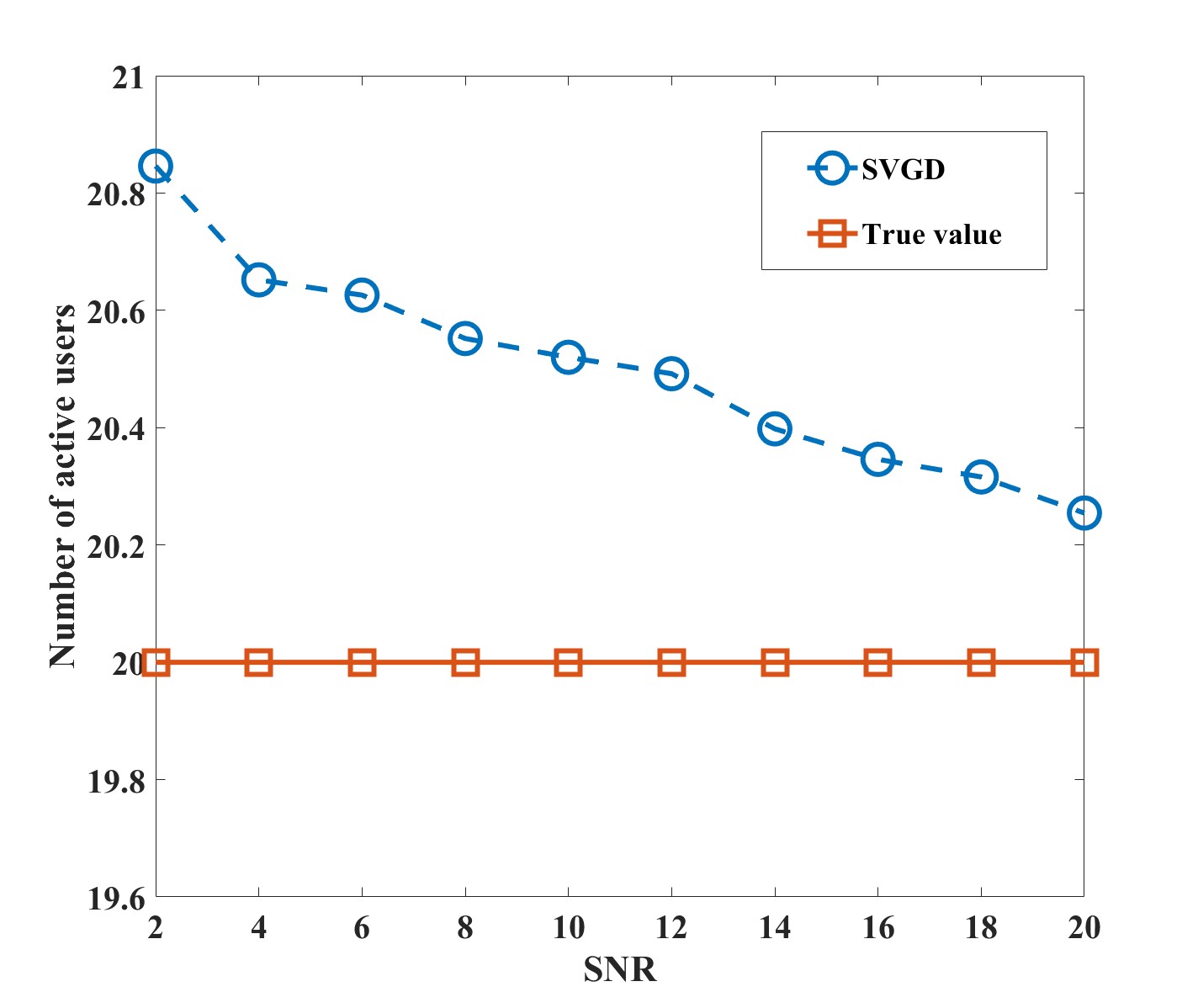}
	\caption{Estimation of active users in a dense user scenario}
	\label{N}
\end{figure}
Fig. \ref{N} shows the estimation of active users in a dense user scenario with $(M, S, N, K)=(20, 10, 20, 30)$. When the signal-to-noise ratio (SNR) is low, there is a large error between the estimated value $\vert\vert \tilde{\mathbf{x}} \vert\vert_{1}$ and the true value $\vert\vert {\mathbf{x}} \vert\vert_{1}$. With SNR increasing, $\bigtriangledown_{\mathbf{x}_i^{t}}\log q(\mathbf x_i^{t})$ becomes non-negligible. Therefore, the SGVD detector can extract effective gradient information to update particles, which improves the detection accuracy.

\begin{algorithm}[ht]
	\caption{Normalized SVGD Detector with Momentum}
        \label{alg: optimized SVGD}
	\LinesNumbered 
	\KwIn{A target function $q(\mathbf x)$ and  a set of initial particles $\{\mathbf x_{i}^0\}_{i=1}^n$}
	\KwOut{A set of particles $\{\tilde{\mathbf x}_{i}\}_{i=1}^n$ that approximates the target function}
	\For{iteration $t$}{
        Calculate the error $\Delta$ using Eq. (\ref{Eq:error}).
        
        Update the gradient ${\varphi}(\mathbf x_i^{t})$ using Eq. (\ref{Eq:gradient}).

        \tcc{Accumulate history gradients}
        \If {$\alpha \neq 0$}               
        {
            \uIf {$t > 1$} {
                ${\mathbf g}_t\leftarrow \alpha {\mathbf g}_{t-1}+(1-\alpha){\varphi}^2(\mathbf x_i^{t})$
            } \Else {${\mathbf g}_t\leftarrow {\varphi}^2(\mathbf x_i^{t})$} 
        }                

        Compute ${\varphi}(\mathbf x_i^{t})$ using Eq. (\ref{Eq:accumulate the history gradients}).
        
        \tcc{Weight Decay}
        \If{$\lambda \neq 0$}
        {
                ${\varphi}(\mathbf x_i^{t})\leftarrow {\varphi}(\mathbf x_i^{t})-\lambda \mathbf x_i^{t} $
        }

        \tcc{Gradient with Momentum}
        \If {$\gamma \neq 0$} 
        {
            \uIf {$t > 1$} {
                ${\mathbf r}_t\leftarrow \gamma {\mathbf r}_{t-1}+(1-\gamma){\varphi}(\mathbf x_i^{t})$
            } \Else {${\mathbf r}_t\leftarrow {\varphi}(\mathbf x_i^{t})$} 
            $ {\varphi}(\mathbf x_i^{t})\leftarrow {\mathbf r}_t$
        }                       

		Compute $\mathbf x_i^{t+1}$ using Eq. (\ref{Eq:final_step}).
	}
\end{algorithm}
\subsection{Normalized SVGD Detector with Momentum}

The irregular gradients and noise have a negative influence on the update of particles. To overcome these limitations of the SVGD detector, we propose a normalized SVGD (NSVGD) detector with momentum. Firstly, as is shown in Algorithm \ref{alg: optimized SVGD}, a bias correction term $\Delta$ is added to the updating process to improve the anti-interference performance. This bias term $\Delta$ calculates the error between the estimated and actual number of active users using  
\begin{equation}\label{Eq:error}
    \Delta \leftarrow \mu \left(nN-\sum_{i=1}^{n}\vert\vert\mathbf{x}_{i}^{t}\vert\vert_{1}\right).
\end{equation}
This error corrects the updating direction of particles. In a noisy environment, although $\bigtriangledown_{\mathbf{x}_i^{t}}\log q(\mathbf x_i^{t})$ will disappear during the update process, the bias term $\Delta$ can still provide useful information for the update of particles. Therefore, the bias term $\Delta$ can improve the robustness of the detector in different environments.

Next, we compute the gradient using Eq.~(\ref{Eq:gradient}).
Then, we accumulate the history gradients and normalize the accumulated gradient:
\begin{equation}\label{Eq:accumulate the history gradients}
 {\varphi}(\mathbf x_i^{t})\leftarrow \frac{{\varphi}(\mathbf x_i^{t})}{\epsilon+\sqrt{{\mathbf g}_t}},
 \end{equation}
where $\sqrt{(\cdot)}$ represents element-wise square root.
These operations solve the problem of continuous decline in learning rate in comparison to the AdaGrad method~\cite{ruder2016overview}. Additionally, the weight decay~\cite{loshchilov2018fixing} is applied to adjust the gradient to find an optimal result. 
Moreover, we apply momentum, a strategy that assists in navigating high error and low curvature regions~\cite{atici2022normalized}. Finally, the particles are updated according to the gradient and estimated error:
\begin{equation}\label{Eq:final_step}
		\mathbf x_i^{t+1}\leftarrow \mathbf x_i^{t}+\varepsilon{\varphi}(\mathbf x_i^{t})+\Delta.
 \end{equation}

\section{Experimental Results}\label{simulation}
In this section, we compare our methods with the MCMC method using Gibbs sampling~\cite{choi2018mcmc}. We consider non-orthogonal preambles. The elements of $\mathbf{P}$ are independent CSCG random variables with zero-mean and variance $\frac{1}{S}$, \textit{i.e.}, $[{\rm {P}}]_{u,v}\sim \mathcal{CN}(0, \frac{1}{S})$. 
We initialize vector $\mathbf x $ in a uniform distribution on $[1, 1.1]$ in both the SVGD detector and the normalized SVGD detector with momentum. In addition, we use Gaussian radial basis function (RBF) kernel~\cite{kuo2013kernel}. It is defined as $k(\mathbf x,\mathbf x')=\exp(-\frac{1}{h} \Vert \mathbf x-\mathbf x' \Vert_{2}^2)$, where bandwidth $h=\frac{\rm{med}^2}{\log n}$, and med is the median of the pairwise distance between the current points $\{\mathbf x_i\}^n_{i=1}$. The parameters $n$, $\mu$, $\alpha$, ${\epsilon}$, $\gamma$, $\varepsilon$, and $\lambda$ are chosen as 6, 0.01, 0.9, 1, 0.9, 0.01, and 0.1, respectively. We use $N_{iteration}$ to represent the number of iterations to obtain stable particles. The sample mean can be obtained by all particles: 
\begin{equation}
\bar{\mathbf x}=\frac{1}{n}\sum_{i=1}^{n}\tilde{\mathbf x}_{i}=[\bar{x}_0,\dots,\bar{x}_m,\dots,\bar{x}_M].
\end{equation}
We use the rounded sample mean to estimate $x_m$, $\textit{i.e.}$, $\hat x_m=\lfloor\bar{x}_m\rceil\in\{0,\dots,N\}$, where $\lfloor x\rceil$ represents the nearest integer of $x$.

We consider two performance measures. The first one is the mean square error (MSE).
The second one is to test the accuracy of different methods named the probability of activity detection error: 
	\begin{eqnarray}\label{KKK}
		\begin{array}{l}
			{\rm{P_{ADE}}}={\rm {Pr}}(x_m\neq \hat{x}_m).
		\end{array}
	\end{eqnarray}

\begin{figure}[ht]
	\centering
	\includegraphics[scale=0.265]{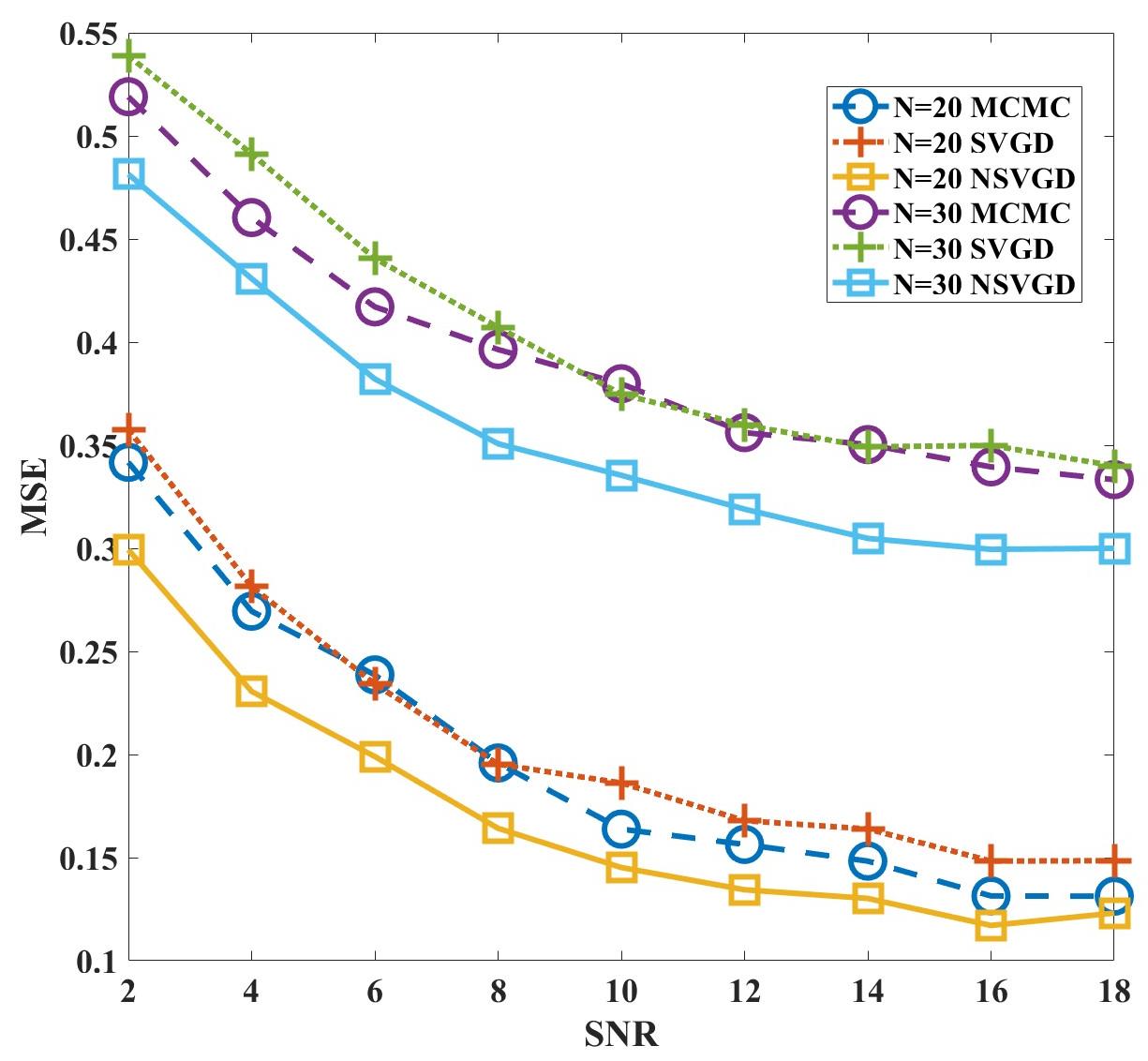}\\
	\caption{MSE for different SNR when $M=20, S=10, K=30$.}
        \label{Fig: MSE}
\end{figure}

\begin{figure}[ht]
	\centering
	\includegraphics[scale=0.25]{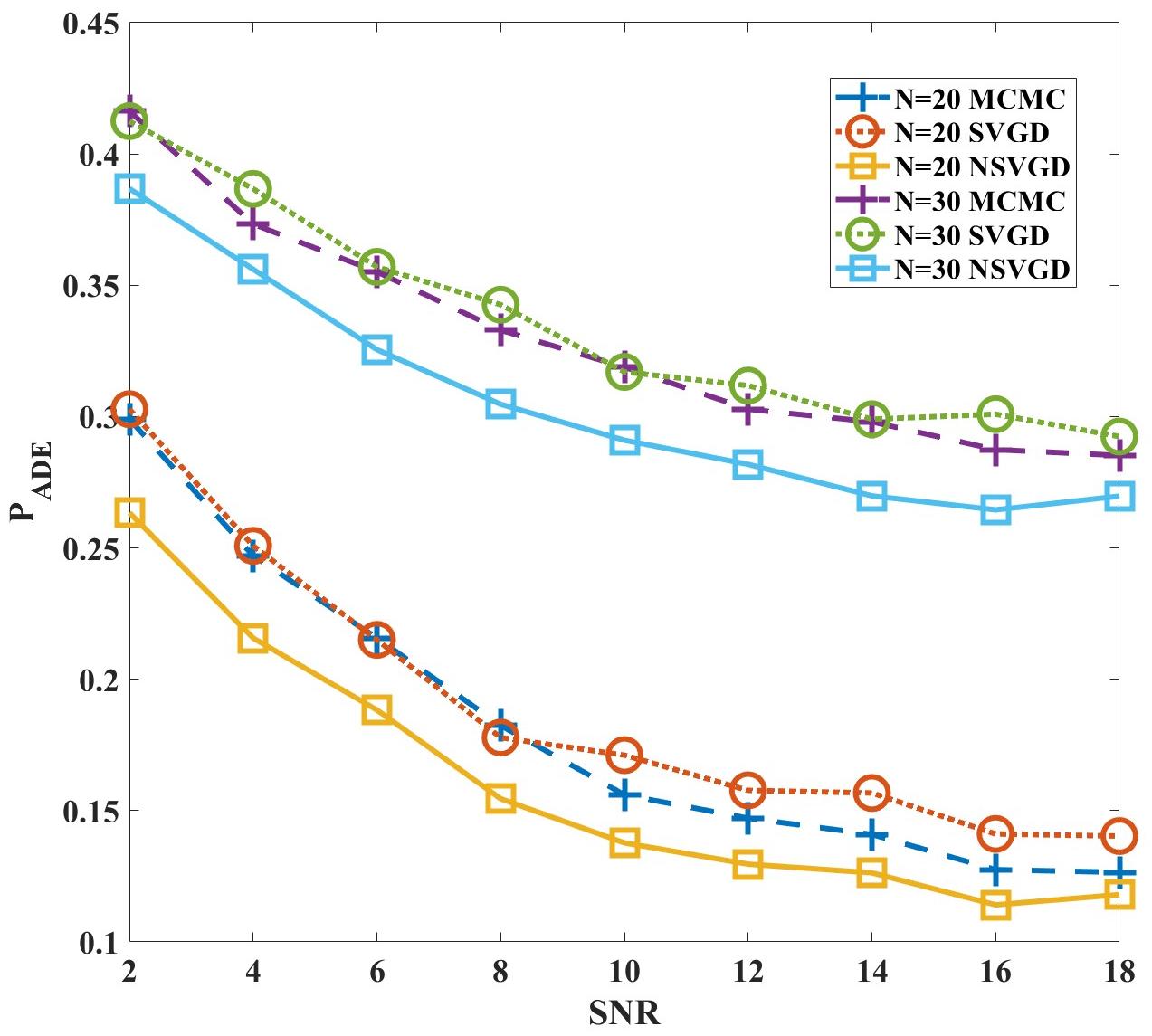}\\
	\caption{Probability of ADE for different SNR when $M=20, S=10, K=30$.}
        \label{Fig: P}
\end{figure}

\begin{figure}[!ht]
	\centering
	\includegraphics[scale=0.265]{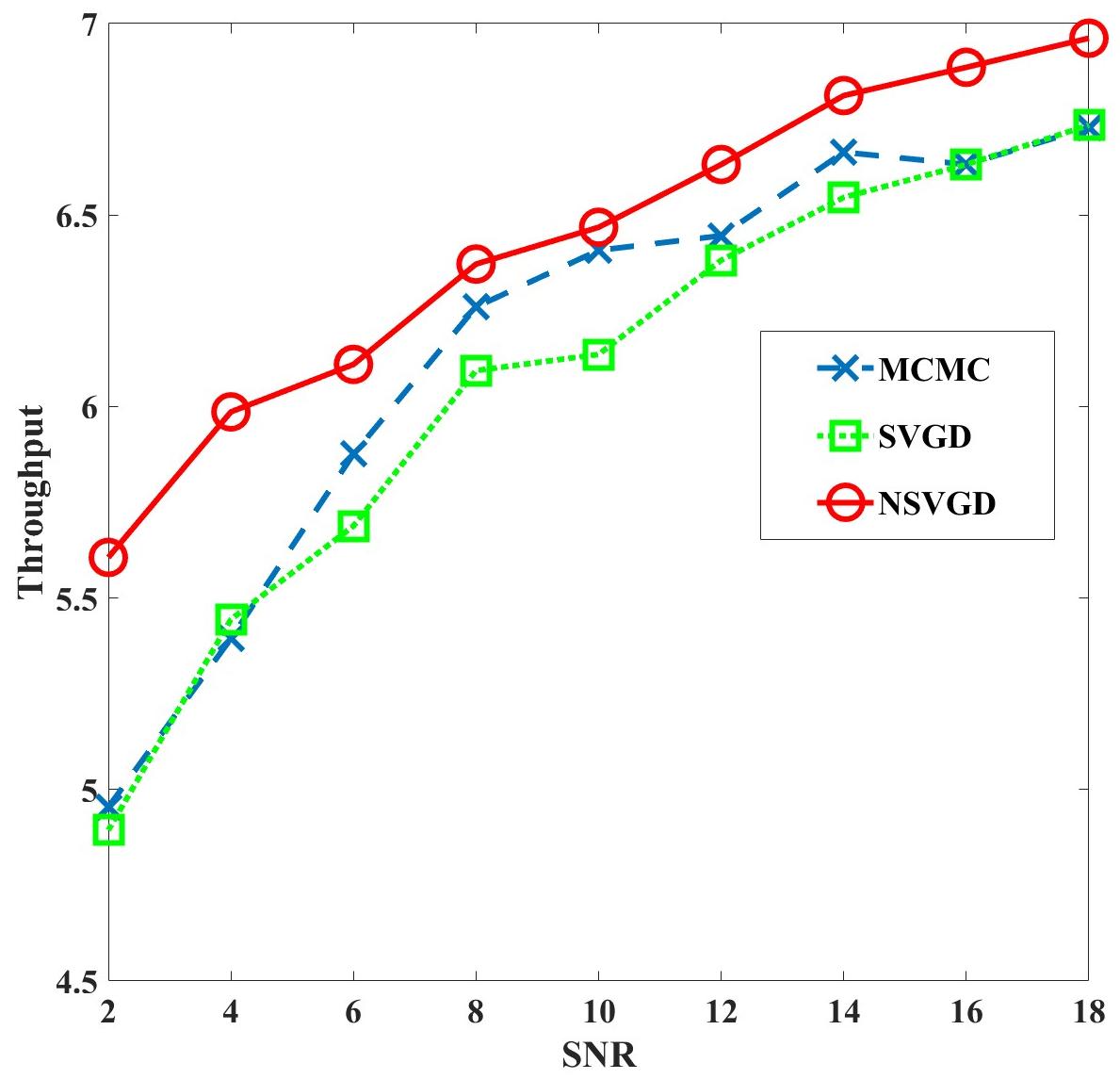}\\
	\caption{Throughput for different SNR when $M=20, S=10, K=30, N=20$.}
        \label{Fig: Th}
\end{figure}

In Fig.~\ref{Fig: MSE} and Fig.~\ref{Fig: P}, with $\rm{P_{ADE}}$ and MSE, we compare the performance of the NSVGD detector with momentum, the SVGD detector, and the MCMC-based detector for different values of SNR in a dense user scenario. The NSVGD detector with momentum outperforms the SVGD detector and the MCMC approach. The improvement of performance at a low SNR shows that the NSVGD detector can reduce the impact of external environmental noise on the SVGD detector. The improvement of performance at a high SNR proves that the NSVGD detector has better robustness and self-adaptation than the SVGD detector. In addition, with the number of active users decreasing from 30 to 20, the possibility of preamble collision decreases gradually and the detection accuracy of the three algorithms also improves.

Fig.~\ref{Fig: Th} shows the results of throughput. Throughput represents the number of preambles without false detection and collision. $M = 20$, $S =10$, $N =20$, $K =30$. It is observed that the NSVGD detector achieves a higher throughput than SVGD and MCMC detectors. It indicates that more active users access the base station successfully.

The computation complexities of the MCMC-based detector, SVGD detector and NSVGD detector are $O(MS(S + K))$, $O(nMKS^2)$ and $O(nMKS^2)$ respectively. It is noticed that the computational complexities of the three detectors are all linearly proportional to the number of preambles $M$ and do not change when the number of active users increases. Therefore, three detectors have a similar computation complexity.

\section{Conclusion}\label{conclusion}
In this paper, we propose a novel preamble detection algorithm based on SVGD, which efficiently leverages deterministic updates of particles to complete preamble detection. Through the error analysis, the performance of the SVGD detector degrades due to the noise. To solve this problem, we propose the NSVGD detector with momentum, which adds a bias correction term to enhance the robustness. Simulation results show that the proposed algorithm performs better than MCMC-based approaches in a dense user scenario. Moreover, the NSVGD detector has a low computation complexity, which is linearly proportional to the number of preambles.
\small
\bibliographystyle{IEEEbib}
\bibliography{strings,refs}

\end{document}